\documentclass[11pt,a4paper]{article}
\usepackage{jheppub}
\usepackage{amsmath}
\usepackage[most]{tcolorbox}
\usepackage{dsfont}
\usepackage{ulem}
\usepackage{natbib}
\usepackage{xcolor}
\usepackage[hang,flushmargin]{footmisc}
\usepackage{tikz-cd}
\usepackage{enumitem}
\usepackage{amsfonts,amssymb, amscd,amsmath,latexsym,amsbsy,bm}
\usepackage{stmaryrd}
\usepackage{todonotes}

\usepackage{float}

\makeatletter\renewcommand{\@biblabel}[1]{#1.}\makeatother

\newtcolorbox{empheqboxed}{colback=gray!20, 
 colframe=white,
 width=\textwidth,
 sharpish corners,
 top=0mm, 
 bottom=0pt
}

\title{Hyperbolic and trigonometric hypergeometric solutions to the star-star equation}

\author{Erdal Catak$^a$, Ilmar Gahramanov$^{b,c,d}$ and Mustafa Mullahasanoglu$^{b}$}
\affiliation{
$^{a}$ Department of Physics, Istanbul University,\\ 34134 Istanbul, Turkey \\[-0.4cm]

$^b$ Department of Physics, Bogazici University,\\ 34342 Bebek, Istanbul, Turkey\\[-0.4cm]

$^c$ Department of Mathematics, Khazar University, \\ Mehseti St. 41, AZ1096, Baku, Azerbaijan\\[-0.4cm]

$^{d}$ Institute of Radiation Problems, Azerbaijan National Academy of Sciences, \\ B.Vahabzade St. 9, AZ1143, Baku, Azerbaijan\\[-0.2cm]

}

\emailAdd{ecatak@istanbul.edu.tr}
\emailAdd{ilmar.gahramanov@boun.edu.tr}
\emailAdd{mustafa.mullahasanoglu@boun.edu.tr}

\abstract{We construct the hyperbolic and trigonometric solutions to the star-star relation  via the gauge/YBE correspondence by using the three-dimensional lens partition function and superconformal index for a certain  $\mathcal N=2$ supersymmetric gauge dual theories. This correspondence relates supersymmetric gauge theories to exactly solvable models of statistical mechanics. The equality of partition functions for the three-dimensional supersymmetric dual theories can be written as an integral identity for hyperbolic and basic hypergeometric functions. 
}

\keywords{Star-star relation, star-triangle relation, integrable lattice spin model, gauge/YBE correspondence.}

\begin{document}
\maketitle
\flushbottom

\section{Introduction}

The gauge/YBE correspondence \cite{Spiridonov:2010em,Yamazaki:2012cp} connecting supersymmetric gauge theories and integrable lattice models of statistical mechanics provides a powerful tool for studying spin models. It turns out that most of integrable edge-interacting (Ising-like) models in statistical mechanics \cite{Gahramanov:2015cva,Kels:2015bda,Kels:2017vbc} (and some IRF models \cite{Yamazaki:2013nra,Yamazaki:2015voa,Gahramanov:2017idz}) can be obtained by this correspondence. We will start with a very short account of this topic, the interested reader can find an exhaustive review on the subject in \cite{Gahramanov:2017ysd,Yamazaki:2018xbx}, where necessary information about the correspondence presented. Similar identities appear in integrable models of statistical mechanics. In this work, we present some new hypergeometric integral identities of hyperbolic and trigonometric types.

One of the striking features of recent developments in non-perturbative supersymmetric gauge theories is their deep relationship with interesting mathematical structures, see, e.g. \cite{Gahramanov:2015tta, Spiridonov:2019kto, Tachikawa:2017byo}. At present, they provide the main source of many new identities for hypergeometric functions \cite{Spiridonov:2009za,Spiridonov2014,Krattenthaler:2011da, Gahramanov:gka,Dolan:2011rp}.


In a recent work \cite{Bozkurt:2020gyy}, the authors constructed a new solution to the star-triangle equation. This was achieved by using a certain three-dimensional supersymmetric dual theories on the lens space $S_b^3/\mathbb{Z}_r$. 
The sufficient condition for the integrability of the lattice spin models is the star-star relation  \cite{Baxter:1997tn}. In this paper we present the corresponding star-star relation for the model studied in \cite{Bozkurt:2020gyy} and also for models discussed in \cite{Bazhanov:2007vg,Gahramanov:2013rda,Gahramanov:2014ona} (the corresponding gauge theories live on the squashed sphere $S_b^3$ and $S^2 \times S^1$). In the context of the gauge/YBE correspondence, this relation can be obtained from the equality of three-dimensional $\mathcal N=2$ supersymmetric  partition functions for a certain dual SQED theories. Our first two solutions to the star-star relation are given in terms of hyperbolic hypergeometric integrals (they are written in terms of hyperbolic gamma functions) and the last solution is a trigonometric type written in terms of basic hypergeometric integrals. 


The paper is organized as follows. In Section \ref{IRF} we briefly recall the star-star relation for the IRF models. Then we present solutions to the star-star equation.

\section{Star-star relation}\label{IRF}

We deal here with the integrable interaction-round-a-face (IRF) lattice spin models \cite{baxter1980fundamental,baxter2016exactly}. In the IRF models, spin variables are located on the sites of the square lattice and interact via face. The sufficient condition for the integrability is the Yang-Baxter equation. In \cite{baxter:1997ssr} it was shown that the Yang-Baxter equation for the face models can be reduced to study of the so-called star-star relation. The star-star relation contains essentially all the information needed to solve the lattice spin model. Here we mainly follow the work of Baxter\footnote{We use the same notation as in \cite{Bozkurt:2020gyy}.} \cite{baxter:1997ssr}, therefore we refer the reader to the original paper for details.


We consider an IRF model with the face Boltzmann weight
\begin{align}
    R\left(\begin{array}{cc}
    \sigma_4     & \sigma_3 \\
\sigma_1         & \sigma_2
    \end{array}\right)=\sum_{m_i}\int dx_i W(\sigma_1,\sigma_i)W(\sigma_2,\sigma_i)W(\sigma_3,\sigma_i)W(\sigma_4,\sigma_i) \;, 
\end{align}
where the $\sigma_i=(x_i; m_i)$ stands for the  continuous valued spin $x_i$ and discrete valued spin $m_i$ and $W(\sigma_i, \sigma_j)$ denotes the interaction between spins $\sigma_i$ and $\sigma_j$. The Boltzmann weight $W(\sigma_i, \sigma_j)$ solves the star-triangle relation for a certain integrable Ising-type lattice model
\begin{align}\nonumber
   \sum_{m_i} \int dx_i& W_{\alpha_1}(\sigma_1,\sigma_i)W_{\alpha_2}(\sigma_2,\sigma_i)W_{\alpha_3}(\sigma_3,\sigma_i) \\ &=\mathcal{R}(\alpha_1,\alpha_2,\alpha_3) W_{\eta-\alpha_1}(\sigma_1,\sigma_2)W_{\eta-\alpha_2}(\sigma_1,\sigma_3)W_{\eta-\alpha_3}(\sigma_2,\sigma_3) \;,
\end{align}
where $\alpha_1+\alpha_2+\alpha_3=\eta$.
Here $\alpha_i$ stands for the rapidity parameter (spectral parameter) and $\mathcal R$ is a spin-independent function. 

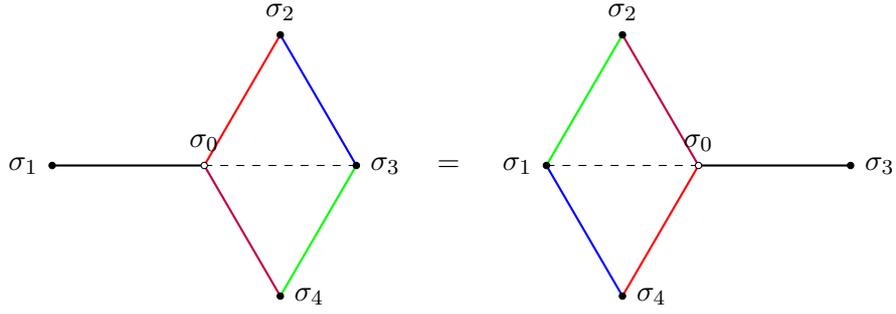
\begin{figure}[t!]
\centering

\begin{tikzpicture}[scale=1]

\draw[-,black,thick] (5.5,0)--(7.5,0);
\draw[-,red,thick] (8.5,1.732)--(7.5,0);
\draw[-,green,thick] (9.5,0)--(8.5,-1.732);
\draw[-,dashed] (7.5,0)--(9.5,0);
\draw[-,blue,thick] (9.5,0)--(8.5,1.732);
\draw[-,purple,thick] (7.5,0)--(8.5,-1.732);

\filldraw[fill=black,draw=black] (5.5,0) circle (1.2pt)
node[left=1.5pt] {\color{black} $\sigma_1$};
\filldraw[fill=black,draw=black] (9.5,0) circle (1.2pt)
node[right=1.5pt] {\color{black} $\sigma_3$};
\filldraw[fill=black,draw=black] (8.5,1.732) circle (1.2pt)
node[above=1.5pt] {\color{black} $\sigma_2$};
\filldraw[fill=black,draw=black] (8.5,-1.732) circle (1.2pt)
node[right=1.5pt] {\color{black} $\sigma_4$};
\filldraw[fill=white,draw=black] (7.5,0) circle (1.2pt)
node[above=1.5pt] {\color{black} $\sigma_0$};

\fill[white!] (11,0) circle (0.01pt)
node[left=0.05pt] {\color{black}$=$};

\draw[-,dashed] (12,0)--(14,0);
\draw[-,purple,thick] (13,1.732)--(14,0);
\draw[-,blue,thick] (12,0)--(13,-1.732);
\draw[-,black,thick] (14,0)--(16,0);
\draw[-,green,thick] (12,0)--(13,1.732);
\draw[-,red,thick] (14,0)--(13,-1.732);

\filldraw[fill=black,draw=black] (12,0) circle (1.2pt)
node[left=1.5pt] {\color{black} $\sigma_1$};
\filldraw[fill=black,draw=black] (16,0) circle (1.2pt)
node[right=1.5pt] {\color{black} $\sigma_3$};
\filldraw[fill=black,draw=black] (13,1.732) circle (1.2pt)
node[above=1.5pt] {\color{black} $\sigma_2$};
\filldraw[fill=black,draw=black] (13,-1.732) circle (1.2pt)
node[right=1.5pt] {\color{black} $\sigma_4$};
\filldraw[fill=white,draw=black] (14,0) circle (1.2pt)
node[above=1.5pt] {\color{black} $\sigma_0$};

\end{tikzpicture}
\caption{The star-star relation.}
\end{figure}

Now let us define the following Boltzmann weight
\begin{align}
    \textcolor{red}{R\left(\begin{array}{ccc}
       & \sigma_1 &\\
\sigma_2         & & \sigma_3  \\
&\sigma_4  &
    \end{array}\right)}=W(\sigma_1,\sigma_3)W(\sigma_2,\sigma_4)R\left(\begin{array}{cc}
    \sigma_4     & \sigma_3 \\
\sigma_1         & \sigma_2
    \end{array}\right) \;.\label{R4}
\end{align}
The system is integrable if the Boltzmann weight satisfies the following star-star relation
\begin{align}
    R_{(left)}\left(\begin{array}{ccc}
       & \sigma_1 &\\
\sigma_2         & & \sigma_3  \\
&\sigma_4  &
    \end{array}\right)=R_{(right)}\left(\begin{array}{ccc}
       & \sigma_1 &\\
\sigma_2         & & \sigma_3  \\
&\sigma_4  &
    \end{array}\right) \:.\label{sseq}
\end{align}
The identity (\ref{sseq}) is illustrated in Fig.1, where we skipped the rapidity lines for convenience, for the full picture and explicit expressions the reader is referred to \cite{Baxter:1997tn}.

One can also see (\ref{sseq}) as the following way with the definition (\ref{R4}) to be convinced with the pictorial representation of the star-star relation in Fig.1
\begin{align}
R\left(\begin{array}{cc}
    \sigma_4     & \sigma_3 \\
\sigma_1         & \sigma_2
    \end{array}\right)
=\frac{W(\sigma_1,\sigma_2)W(\sigma_1,\sigma_4)}{W(\sigma_3,\sigma_2)W(\sigma_3,\sigma_4)}
R\left(\begin{array}{cc}
    \sigma_4    & \sigma_3 \\
\sigma_1         & \sigma_2
    \end{array}\right) \:.
\end{align}

where $R\left(\begin{array}{cc}
    \sigma_4    & \sigma_3 \\
\sigma_1         & \sigma_2
    \end{array}\right)$ functions differ with the spectral parameters \cite{Baxter:1997tn} omitted in this study.

By using the star-star relation one obtains the following IRF Yang-Baxter equation (it is depicted in Fig.2)
\begin{align} \nonumber
\sum_{m_i\in \mathbb{Z}}\int d x_i ~~
    \textcolor{red}{R\left(\begin{array}{ccc}
        & \sigma_5 & \\
\sigma_6         &  & \sigma_i \\
&              \sigma_1     &
    \end{array}\right)}  
    \textcolor{blue}{R\left(\begin{array}{cc}
    \sigma_i     & \sigma_3 \\
\sigma_1         & \sigma_2
    \end{array}\right)}    \textcolor{green}{R\left(\begin{array}{cc}
    \sigma_5     & \sigma_4 \\
\sigma_i         & \sigma_3
    \end{array}\right)}  ~~~~~~~~~~~~~~~~~~~~~~\\ 
    ~~~~~~~~~~~~~~~~~~~~~~~~~~~=
    \sum_{m_j\in \mathbb{Z}} \int dx_j ~~
        \textcolor{blue}{R\left(\begin{array}{cc}
    \sigma_6     & \sigma_j \\
\sigma_1         & \sigma_2
    \end{array}\right)}    \textcolor{green}{R\left(\begin{array}{cc}
    \sigma_5     & \sigma_4 \\
\sigma_6         & \sigma_j
    \end{array}\right)}    \textcolor{red}{R\left(\begin{array}{ccc}
        & \sigma_4 & \\
\sigma_j         &  & \sigma_3 \\
&              \sigma_2     &
    \end{array}\right)}\:,
\end{align}
where the summation and integration stand for the discrete and continuous spin variables, respectively. Note that there are several solutions to the IRF Yang-Baxter equation obtained via gauge/YBE correspondence \cite{Yamazaki:2012cp,Gahramanov:2017idz,Gahramanov:2015cva,Yamazaki:2015voa,Kels:2017vbc}.

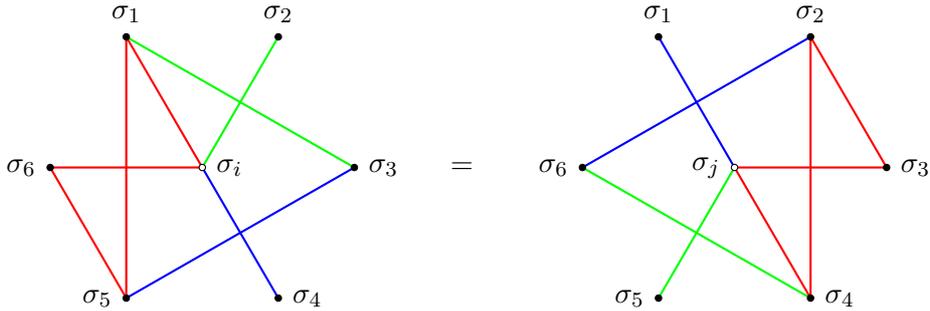
\begin{figure}[t!]
\centering
    
\begin{tikzpicture}[scale=1]

\draw[-,blue,thick] (9,1.732)--(9.5,0.866);
\draw[-,blue,thick] (8,0)--(9.5,0.866);
\draw[-,red,thick] (11,1.732)--(11,0);
\draw[-,blue,thick] (9.5,0.866)--(10,0);
\draw[-,blue,thick] (9.5,0.866)--(11,1.732);
\draw[-,red,thick] (11,0)--(12,0);
\draw[-,red,thick] (10,0)--(11,0);
\draw[-,red,thick] (11,0)--(11,-1.732);
\draw[-,green,thick] (10,0)--(9.5,-0.866);
\draw[-,green,thick] (11,-1.732)--(9.5,-0.866);
\draw[-,green,thick] (9,-1.732)--(9.5,-0.866);
\draw[-,green,thick] (8,0)--(9.5,-0.866);
\draw[-,red,thick] (10,0)--(11,-1.732);
\draw[-,red,thick] (11,1.732)--(12,0);

\filldraw[fill=black,draw=black] (9,1.732) circle (1.2pt)
node[above=1.5pt] {\color{black} $\sigma_1$};
\filldraw[fill=black,draw=black] (8,0) circle (1.2pt)
node[left=1.5pt] {\color{black} $\sigma_6$};
\filldraw[fill=black,draw=black] (12,0) circle (1.2pt)
node[right=1.5pt] {\color{black} $\sigma_3$};
\filldraw[fill=black,draw=black] (11,1.732) circle (1.2pt)
node[above=1.5pt] {\color{black} $\sigma_2$};
\filldraw[fill=black,draw=black] (9,-1.732) circle (1.2pt)
node[left=1.5pt] {\color{black} $\sigma_5$};
\filldraw[fill=black,draw=black] (11,-1.732) circle (1.2pt)
node[right=1.5pt] {\color{black} $\sigma_4$};

\filldraw[fill=white,draw=black] (10,0) circle (1.2pt)
node[left=1.5pt] {\color{black} $\sigma_j$};

\fill[white!] (6.7,0) circle (0.01pt)
node[left=0.05pt] {\color{black}$=$};

\draw[-,red,thick] (2,1.732)--(2,0);
\draw[-,red,thick] (1,0)--(2,0);
\draw[-,green,thick] (4,1.732)--(3.5,0.866);
\draw[-,red,thick] (2,0)--(3,0);
\draw[-,red,thick] (1,0)--(2,-1.732);
\draw[-,red,thick] (2,1.732)--(3,0);
\draw[-,green,thick] (3,0)--(3.5,0.866);
\draw[-,blue,thick] (3,0)--(3.5,-0.866);
\draw[-,blue,thick] (4,-1.732)--(3.5,-0.866);
\draw[-,blue,thick] (2,-1.732)--(3.5,-0.866);
\draw[-,blue,thick] (5,0)--(3.5,-0.866);
\draw[-,green,thick] (5,0)--(3.5,0.866);
\draw[-,green,thick] (3.5,0.866)--(2,1.732);
\draw[-,red,thick] (2,-1.732)--(2,0);

\filldraw[fill=black,draw=black] (2,1.732) circle (1.2pt)
node[above=1.5pt] {\color{black} $\sigma_1$};
\filldraw[fill=black,draw=black] (1,0) circle (1.2pt)
node[left=1.5pt] {\color{black} $\sigma_6$};
\filldraw[fill=black,draw=black] (5,0) circle (1.2pt)
node[right=1.5pt] {\color{black} $\sigma_3$};
\filldraw[fill=black,draw=black] (4,1.732) circle (1.2pt)
node[above=1.5pt] {\color{black} $\sigma_2$};
\filldraw[fill=black,draw=black] (2,-1.732) circle (1.2pt)
node[left=1.5pt] {\color{black} $\sigma_5$};
\filldraw[fill=black,draw=black] (4,-1.732) circle (1.2pt)
node[right=1.5pt] {\color{black} $\sigma_4$};

\filldraw[fill=white,draw=black] (3,0) circle (1.2pt)
node[right=1.5pt] {\color{black} $\sigma_i$};

\end{tikzpicture}
\caption{Yang-Baxter relation via the star-star relation. Spin variables live on vertices and interact via edges.}
\end{figure}

\section{Solutions to the star-star equation}\label{Hypesol}

By using the gauge/YBE correspondence one can systematically derive solution of the Yang-Baxter equation from calculations of supersymmetric gauge theory. In the context of this correspondence the Yang-Baxter equation expresses the identity of partition functions for supersymmetric dual pairs. Therefore the main step is to choose appropriate supersymmetric duality. Here we consider the following three-dimensional $\mathcal N=2$ dual theories \cite{Gahramanov:2014ona,Gahramanov:2016wxi,Bozkurt:2020gyy}
\begin{itemize}
    \item \textbf{theory A} has $U(1)$ gauge group,  six chiral multiplets with $SU(3) \times SU(3) \times U(1)$ global symmetry group 
    \item \textbf{theory B} consists of nine free ``mesons'' with the same global symmetry group as theory A.
\end{itemize}

The supersymmetric localization technique\footnote{The short review of the three-dimensional supersymmetric localization can be found in \cite{Hosomichi:2014hja,Willett:2016adv}.} \cite{Pestun:2007rz} enables us to calculate the partition function on different manifolds. The results of Coulomb branch localization on $S_b^3$ , $S_b^3/{\mathbb Z}_r$ and $S^2 \times S^1$ are known (see, e.g. \cite{Kapustin:2010xq,Imamura:2012rq,Imamura:2011su}) and  we will use these results in order to construct the star-triangle relation and corresponding star-star relation in the next sections. 

\subsection{Notations}

Let us introduce some definitions and notations of special functions which we use in the paper. The $q$-Pochhammer symbol is defined as follows 
\begin{align}
    (z;q)_{\infty}=\prod_{i=0}^{\infty}(1-zq^i)\;.
\end{align}
We use the shorthand notation
\begin{align}
    (z,x;q)_\infty=(z;q)_\infty(x;q)_\infty \;.
\end{align}

We also use hyperbolic gamma function which can be defined   as
\begin{align}
    \gamma^{(2)}(z;\omega_{1},\omega_{2})=e^{\frac{\pi i}{2}B_{2,2}(z;\omega_{1},\omega_{2})}\frac{(e^{-2\pi i\frac{z}{\omega_{2}}}\tilde{q};\tilde{q})_\infty}{(e^{-2\pi i\frac{z}{\omega_{1}}};q)_\infty} \; ,
\end{align}
with the parameters $\tilde{q}=e^{2\pi i \omega_{1}/\omega_{2}}$ and $q=e^{-2\pi i \omega_{2}/\omega_{1}}$ and the $B_{2,2} (z; \omega_1,\omega_2)$ stands for the Bernoulli polynomial 
\begin{align}
    B_{2,2}(z;\omega_{1},\omega_{2})=\frac{z^2-z(\omega_{1}+\omega_{2})}{\omega_{1}\omega_{2}}+\frac{\omega_{1}^2+3\omega_{1}\omega_{2}+\omega_{2}^2}{6\omega_{1}\omega_{2}}\;.
\end{align}

The hyperbolic gamma function also has an integral representation \footnote{One can find different integral representations in \cite{Faddeev:1995nb,woronowicz2000quantum}.} \begin{align}
    \gamma^{(2)}(z;\omega_{1},\omega_{2})=\exp{\left(-\int_{0}^{\infty}\frac{dx}{x}\left[\frac{\sinh{x(2z-\omega_{1}-\omega_{2})}}{2\sinh{(x\omega_{1})}\sinh{(x\omega_{2})}}-\frac{2z-\omega_{1}-\omega_{2}}{2x\omega_{1}\omega_{2}}\right]\right)} \; ,
\end{align}
where $Re(\omega_{1}),Re(\omega_{2})>0$ and $Re(\omega_{1}+\omega_{2})>Re(z)>0$. 

Additionally, we will use  the following reflection property for hyperbolic gamma function
\begin{align} \label{reflectionp}
    \gamma^{(2)}(\omega_1+\omega_2 - z;\omega_1,\omega_2)\gamma^{(2)}(z;\omega_1,\omega_2) \ = \ 1 \;.
\end{align}

\subsection{Solution via \texorpdfstring{$S_b^3$}{Sb3} supersymmetric partition function}

The equivalence of the partition functions for dual theories on $S_b^3$ gives the following hyperbolic hypergeometric integral identity \cite{Kashaev:2012cz,Spiridonov:2010em,BultThesis}

\begin{align} 
\int_{-\textup{i} \infty}^{\textup{i} \infty} \prod_{i=1}^3 \gamma^{(2)}(a_i - x;\omega_1,\omega_2) \gamma^{(2)}(b_i + x;\omega_1,\omega_2) \frac{dx}{\textup{i} \sqrt{\omega_1\omega_2}} = \prod_{i,j=1}^3 \gamma^{(2)}(a_i+b_j;\omega_1,\omega_2)\label{Sb3} \;,
\end{align}
with the balancing condition $\sum_{i=1}^3 (a_i+b_i) = \omega_1+\omega_2$. In \cite{Bazhanov:2007mh,Bazhanov:2007vg} it was shown that this integral gives the star-triangle relation for the Faddeev-Volkov model which has the following Boltzmann weight
\begin{align}
    W_{\alpha}(x_i,x_j)=&\gamma^{(2)}(-\alpha+x_i-x_j;\omega_{1},\omega_{2})
    \gamma^{(2)}(-\alpha-x_i+x_j;\omega_1,\omega_{2}) \;,
\end{align}
where we introduced new variables  $ a_i  =-\alpha_i+x_{i} $ and $ b_i=-\alpha_i-x_{i}$. Here $\alpha$ is a spectral parameter, $x_i$ is a spin variable and $\omega_1, \omega_2$ are temperature-like parameters. 

The corresponding star-star relation for this model (for details, see Appendix A) has the following form
\begin{align}
\int_{-\textup{i} \infty}^{\textup{i} \infty}  \prod_{i=1}^4 \gamma^{(2)}(a_i - x;\omega_1,\omega_2) \gamma^{(2)}(b_i + x;\omega_1,\omega_2) \frac{dx}{\textup{i} \sqrt{\omega_1\omega_2}}
\nonumber \\ =
\frac{\prod_{i,j=1}^2 \gamma^{(2)}(a_i+b_j;\omega_1,\omega_2)}{\prod_{i,j=3}^4 \gamma^{(2)}(\tilde{a_i}+\tilde{b_j};\omega_1,\omega_2)}  
\int_{-\textup{i} \infty}^{\textup{i} \infty}  \prod_{i=1}^4 \gamma^{(2)}(\tilde{a_i} - z;\omega_1,\omega_2) \gamma^{(2)}(\tilde{b_i} + z;\omega_1,\omega_2)
\frac{dz}{i\sqrt{\omega_1\omega_2}}\;, \label{Sb3ss}
\end{align} 
with a new balancing condition $\sum_{i=1}^4 (a_i+b_i) =2( \omega_1+\omega_2)$ and we used the following notations
\begin{align}
\begin{aligned}
     \tilde{a}_i & =  a_i+s, & \tilde{b}_i & =  b_i+s, & \text{if} \;\;\; i=1,2\;,
     \\ 
     \tilde{a}_i & =  a_i-s, & \tilde{b}_i & =b_i-s, & \ \text{if} \;\;\; i=3,4\;,
\end{aligned}
\end{align}
with
    \begin{align}
        \begin{aligned}
             s=\frac{1}{2}(\omega_1+\omega_2-a_1-a_2-b_1-b_2)=\frac{1}{2}(-\omega_1-\omega_2+a_3+a_4+b_3+b_4) \;.
        \end{aligned}
    \end{align}
This integral identity was obtained in \cite{van2007properties}. The physical interpretation of this identity discussed in \cite{Dimofte:2012pd}.


\subsection{Solution via $\texorpdfstring {S_b^3/{\mathbb Z}_r}{Sb3Zr}$ supersymmetric partition function}

We again start with the equivalence of the partition functions for dual theories. This time we consider the dual theories on $S_b^3/{\mathbb Z}_r$ and obtain the following hyperbolic hypergeometric integral identity\footnote{Note that this integral identity was obtained via the reduction procedure and it needs to be proven rigorously.} \cite{Bozkurt:2020gyy}
\begin{align}
    \sum_{y=0}^{[ r/2 ]}\epsilon (y) e^{-\pi iy}\int _{-\infty}^{\infty} 
    \prod_{i=1}^3 & \gamma^{(2)}(-i(a_i-x)-i\omega_1(u_i- y);-i\omega_1r,-i\omega) \nonumber \\
     \times & \gamma^{(2)}(-i(a_i-x)-i\omega_2(r-(u_i- y));-i\omega_2r,-i\omega) \nonumber \\
    \times & \gamma^{(2)}(-i(b_i+x)-i\omega_1(v_i+ y);-i\omega_1r,-i\omega) \nonumber
    \\
     \times & \gamma^{(2)}(-i(b_i+x)-i\omega_2(r-(v_i+y));-i\omega_2r,-i\omega)
     \frac{dx}{r\sqrt{-\omega_1\omega_2}}\nonumber\\
     = e^{\frac{-\pi i}{2}\sum_{i=1}^3(u_i-v_i)}\prod_{i,j=1}^3& \gamma^{(2)}(-i(a_i+ b_j)-i\omega_1(u_i+ v_j);-i\omega_1r,-i\omega)\nonumber\\ \times& \gamma^{(2)}(-i(a_i+ b_j)-i\omega_2(r-(u_i+ v_j));-i\omega_2r,-i\omega)) \;, \label{Sb3Zr}
\end{align}
with the balancing conditions $\sum_{i=1}^3a_i+b_i=\omega_1+\omega_2$ and $\sum_{i=1}^3u_i+v_i=0$.  The $\epsilon(y)$ function is defined as $\epsilon(0)=\epsilon(\lfloor\frac{r}{2}\rfloor)=1$ and $\epsilon(y)=2$ otherwise. The main difference from the expression (\ref{Sb3}) is that here the summation is over the holonomies $y=\frac{r}{2\pi} \int A_\mu d x^{\mu}$, where the integration is over a non-trivial cycle on $S_b^3/{\mathbb Z}_r$ and $A_\mu$ is the gauge field, see, e.g. \cite{Imamura:2012rq}. 

By introducing new variables $a_i  =-\alpha_i+x_{i} $ and $ b_i=-\alpha_i-x_{i} $ with the condition $u_i=-v_i$ for $i=\overline{1,3}$,  one can rewrite the integral identity (\ref{Sb3Zr}) as the star-triangle equation with the following Boltzmann weight
\begin{align} \label{Bweight}
    W_{\alpha}(x_i,x_j,u_i,u_j)=&e^{-\pi i(u_i+u_j)}\gamma^{(2)}(-i(-\alpha+x_i-x_j)-i\omega_1(u_i-u_j);-i\omega_{1}r,-i\omega)\nonumber \\&\gamma^{(2)}(-i(-\alpha+x_i-x_j)-i\omega_2(r-(u_i-u_j));-i\omega_2r,-i\omega)\nonumber \\&\gamma^{(2)}(-i(-\alpha-x_i+x_j)-i\omega_1(u_j-u_i);-i\omega_{1}r,-i\omega)\nonumber \\&\gamma^{(2)}(-i(-\alpha-x_i+x_j)-i\omega_2(r-(u_j-u_i));-i\omega_2r,-i\omega) \;.
\end{align}
The model with the Boltzmann weight (\ref{Bweight}) is an exactly solvable lattice spin model with discrete and continuous spin variables living on sites, where $x_i$ represents continuous spin and $u_i$ represents discrete spin variable. The $r=1$ case corresponds to the Faddeev-Volkov model from the previous section\footnote{The $r=2$ case is also special, in \cite{Hadasz:2013bwa} it was used for proving orthogonality and completeness relation of the Clebsch-Gordan coefficients for the self-dual continuous series of $U_q(osp(1|2))$.}.

Using the similar technique presented in Appendix A one can construct the star-star relation for this model
\begin{align}
  \sum_{y=0}^{[ r/2 ]}\epsilon (y) \int _{-\infty}^{\infty} \prod_{i=1}^4  \gamma^{(2)}(-i(a_i-x)-i\omega_1(u_i- y);-i\omega_1r,-i\omega) \nonumber \\
     \times  \gamma^{(2)}(-i(a_i-x)-i\omega_2(r-(u_i- y));-i\omega_2r,-i\omega) \nonumber \\
    \times  \gamma^{(2)}(-i(b_i+x)-i\omega_1(v_i+ y);-i\omega_1r,-i\omega) \nonumber
    \\
     \times  \gamma^{(2)}(-i(b_i+x)-i\omega_2(r-(v_i+y));-i\omega_2r,-i\omega)  
     \frac{dx}{r\sqrt{-\omega_1\omega_2}}\nonumber \\
     =\frac{e^{\frac{\pi i}{2}\sum_{i=1}^2(u_i-v_i)}}{e^{\frac{\pi i}{2}\sum_{i=3}^4(\tilde{u_i}-\tilde{v_i})}}
   \frac{\prod_{i,j=1}^2 \gamma^{(2)}(-i(a_i+ b_j)-i\omega_1(u_i+ v_j);-i\omega_1r,-i\omega) }{\prod_{i,j=3}^4 \gamma^{(2)}(-i(\tilde{a_i}+\tilde{b_j})-i\omega_1(\tilde{u_i}+\tilde{v_j});-i\omega_1r,-i\omega) } 
 \nonumber \\ \times  
     \frac{\prod_{i,j=1}^2  \gamma^{(2)}(-i(a_i+ b_j)-i\omega_2(r-(u_i+ v_j));-i\omega_2r,-i\omega)}{\prod_{i,j=3}^4  \gamma^{(2)}(-i(\tilde{a_i}+\tilde{b_j})-i\omega_2(r-(\tilde{u_i}+\tilde{v_j}));-i\omega_2r,-i\omega) } 
  \nonumber \\ \times  
     \sum_{m=0}^{[ r/2 ]}\epsilon (m) \int _{-\infty}^{\infty} \prod_{i=1}^4  \gamma^{(2)}(-i(\tilde{a_i}-z)-i\omega_1(\tilde{u_i}- m);-i\omega_1r,-i\omega) \nonumber \\
     \times  \gamma^{(2)}(-i(\tilde{a_i}-z)-i\omega_2(r-(\tilde{u_i}- m));-i\omega_2r,-i\omega) \nonumber \\
    \times  \gamma^{(2)}(-i(\tilde{b_i}+z)-i\omega_1(\tilde{v_i}+ m);-i\omega_1r,-i\omega) \nonumber
    \\
     \times  \gamma^{(2)}(-i(\tilde{b_i}+z)-i\omega_2(r-(\tilde{v_i}+m));-i\omega_2r,-i\omega) \frac{dz}{r\sqrt{-\omega_1\omega_2}}  \;,
\end{align}
where the balancing conditions are $\sum_{i=1}^4a_i+b_i=2(\omega_1+\omega_2)$ and $\sum_{i=1}^4u_i+v_i=0$,
and we used the following notations
\begin{align}
\begin{aligned}
     \tilde{a}_i & =  a_i+s, & \tilde{b}_i & =  b_i+s, & \tilde{u}_i & =u_i+p, & \tilde{v}_i  & = v_i+p, & \text{if} \;\;\; i=1,2 \;,
     \\ 
     \tilde{a}_i & =  a_i-s, & \tilde{b}_i & =  b_i-s, & \tilde{u}_i & = u_i-p, & \tilde{v}_i & =v_i-p, & \text{if} \;\;\; i=3,4 \;,
\end{aligned}
\end{align}
where 
\begin{align}
\begin{aligned}
s & =\frac{1}{2}(\omega_1+\omega_2-a_1-a_2-b_1-b_2)=\frac{1}{2}(-\omega_1-\omega_2+a_3+a_4+b_3+b_4) \;,\\
p & =-\frac{1}{2}(u_1+u_2+v_1+v_2)=\frac{1}{2}(u_3+u_4+v_3+v_4) \:.
\end{aligned}
\end{align}


\subsection{Solution via \texorpdfstring{$S^2 \times S^1$}{S2S1} supersymmetric partition function}

In this section we present a new trigonometric solution to the star-triangle equation and to the star-star equation. In \cite{Gahramanov:2013rda,Gahramanov:2014ona,Gahramanov:2016wxi} authors considered the following basic hypergeometric sum/integral identity\footnote{This identity can be written as a pentagon identity which is related to the triangulation of 3-manifolds.} which represents the equivalence of the superconformal indices for dual theories (partition function on $S^2 \times S^1$) discussed in Sec.\ref{Hypesol}
\begin{align}
\sum_{y=-\infty}^\infty\oint \prod_{i=1}^3\frac{(q^{1+(y+u_i)/2}(a_ix)^{-1},q^{1+(v_i-y)/2}xb_i^{-1};q)_\infty}{(q^{(y+u_i)/2}a_ix,q^{(v_i-y)/2}b_ix^{-1};q)_\infty}\frac{1}{x^{3y}}\frac{dx}{2\pi i x} \nonumber \\
=\frac{1}{\prod_{i=1}^3a_i^{u_i}b_i^{v_i}}\prod_{i,j=1}^3\frac{(q^{1+(u_i+v_j)/2}(a_ib_j)^{-1};q)_\infty}{(q^{(u_i+v_j)/2}a_ib_j;q)_\infty} \;,
\end{align}
where the balancing conditions are $\prod_{i=1}^3a_ib_i=q$ and $\sum_{i=1}^3u_i+v_i=0$.

This identity can be written as the star-triangle relation by introducing the new fugacities $a_i=\alpha_i^{-1}x_{i} $ and $b_i=\alpha_i^{-1}x_{i}^{-1}$ and using the condition $u_i=-v_i$. The resulting Boltzmann weight then has the following form
\begin{align}
    W_{\alpha}(x_i,x_j,u_i,u_j)=\frac{(q^{1+(u_i-u_j)/2}(\alpha^{-1}x_ix_j^{-1})^{-1};q)_\infty}{(q^{(u_i-u_j)/2}\alpha^{-1} x_ix_j^{-1};q)_\infty}
    \frac{(q^{1+(u_j-u_i)/2}(\alpha^{-1}x_i^{-1}x_j)^{-1};q)_\infty}{(q^{(u_j-u_i)/2}\alpha^{-1}x_i^{-1}x_j;q)_\infty} \;,
\end{align}
where $\alpha$ stands for the spectral parameter. The corresponding statistical mechanics model is a square lattice model with edge interaction and discrete and continuous spin variables. This model is a special case (with broken gauge symmetry) of the integrable lattice spin model considered in \cite{Gahramanov:2015cva} and it is a trigonometric analogue of the Faddeev-Volkov model.

One we can construct the star-star relation for this model
\begin{align}
\sum_{y=-\infty}^\infty\oint \prod_{i=1}^4\frac{(q^{1+(y+u_i)/2}(a_ix)^{-1},q^{1+(v_i-y)/2}x(b_i)^{-1};q)_\infty}{(q^{(y+u_i)/2}a_ix,q^{(v_i-y)/2}b_jx^{-1};q)_\infty}\frac{1}{x^{5y}}\frac{dx}{2\pi i x}  \nonumber \\
=
\frac{1}{\frac{\prod_{i=1}^2a_i^{u_i}b_i^{v_i}}{\prod_{i=3}^4(\tilde{a_i})^{\tilde{u_i}}(\tilde{b_i})^{\tilde{v_i}}}}
\frac{\prod_{i,j=1}^2\frac{(q^{1+(u_i+v_j)/2}(a_ib_j)^{-1};q)_\infty}{(q^{(u_i+v_j)/2}a_ib_j;q)_\infty}}{\prod_{i,j=3}^4\frac{(q^{1+(\tilde{u_i}+\tilde{v_j})/2}(\tilde{a_i}\tilde{b_j})^{-1};q)_\infty}{(q^{(\tilde{u_i}+\tilde{v_j})/2}\tilde{a_i}\tilde{b_j};q)_\infty}}
\nonumber \\ 
\times \sum_{m=-\infty}^\infty\oint \prod_{i=1}^4\frac{(q^{1+(m+ \tilde{u_i})/2}(\tilde{a_i}z)^{-1},q^{1+(\tilde{v_i}-m)/2}z\tilde{b_i}^{-1};q)_\infty}{(q^{(m+\tilde{u_i})/2}\tilde{a_i}z,q^{(\tilde{v_i}-m)/2}\tilde{b_i}z^{-1};q)_\infty}\frac{1}{z^{5m}}\frac{dz}{2\pi i z} \;, 
\end{align} 
with the new  balancing conditions  $\prod_{i=1}^4a_ib_i=q^2$ and $\sum_{i=1}^4u_i+v_i=0$. In the latter expression we used the following notations
\begin{align}
\begin{aligned}
     \tilde{a}_i & =  a_is, & \tilde{b}_i & =  b_is, & \tilde{u}_i & =u_i+p, & \tilde{v}_i  & = v_i+p, & \text{if} \;\;\; i=1,2\;,
     \\ 
     \tilde{a}_i & =  a_is^{-1}, & \tilde{b}_i & =  b_is^{-1}, & \tilde{u}_i & = u_i-p, & \tilde{v}_i & =v_i-p, & \text{if} \;\;\; i=3,4\;,
\end{aligned}
\end{align}
where 
\begin{align}
\begin{aligned}
s&=\sqrt{\frac{q}{a_1a_2b_1b_2}}=\sqrt{\frac{a_3a_4b_3b_4}{q}}\;,  \\
p & =-\frac{1}{2}(u_1+u_2+v_1+v_2)=\frac{1}{2}(u_3+u_4+v_3+v_4) \;.
\end{aligned}
\end{align}


\section{Conclusions}

In this work, we constructed hyperbolic and trigonometric solutions to the star-star equation. We obtained new solutions from the equality of three-dimensional $\mathcal N=2$ supersymmetric partition functions for certain dual SQED theories via the gauge/YBE correspondence.

There are several ways of constructing solutions to the star-star equation. One can use the Bailey pair construction starting from the star-triangle relation for the models discussed here. It is possible to obtain the solution by breaking the gauge symmetry from $SU(2)$ group to the $U(1)$ for the supersymmetric dual theories with $SU(2)$ gauge group and $SU(6)$ flavor group considered in \cite{Gahramanov:2016ilb}. 

There are many interesting limits of the solutions considered here, for instance it would be interesting to construct solution to the star-star equation in terms of Euler gamma functions \cite{Eren:2019ibl}.

The gauge/YBE correspondence has revealed various interesting connections among integrable models and supersymmetric gauge theories. There are underlying mathematical structures such as quantum algebras related to the solutions of the Yang-Baxter equations obtained via gauge/YBE correspondence. It would be interesting to pursue this direction for these solutions to the star-star relation.


\section*{Acknowledgements}

We are grateful to Deniz N. Bozkurt for valuable discussions. The work of Ilmar Gahramanov is partially supported by the Bogazici University Research Fund under grant number 20B03SUP3 and TUBITAK grant 220N106. Mustafa Mullahasanoglu is supported by the 2209-TUBITAK National/International Research Projects Fellowship Programme for Undergraduate Students under grant number 1919B011902237.

\appendix

\section{Derivation of the star-star relation (\ref{Sb3ss})}

Here we follow the approach presented in \cite{van2007properties}. 

Let us consider the following double integral
\begin{align} 
\int_{-\textup{i} \infty}^{\textup{i} \infty}  \prod_{i=1}^2 \gamma^{(2)}(a_i - x) \gamma^{(2)}(b_i + x)  
 \int_{-\textup{i} \infty}^{\textup{i} \infty}  \prod_{i=3}^4 \gamma^{(2)}(a_i-s - z) \gamma^{(2)}(b_i-s + z)
 \nonumber \\ \times
\gamma^{(2)}(s+z - x) \gamma^{(2)}(s-z + x)\frac{dx}{\textup{i} \sqrt{\omega_1\omega_2}}\frac{dz}{\textup{i} \sqrt{\omega_1\omega_2}} \:. \label{doubleint}
\end{align}
Here we used the shorthand notation $\gamma^{(2)}(x)=\gamma^{(2)}(x;\omega_1, \omega_2)$.
First we integrate the integral (\ref{doubleint}) over the $x$ variable using the identity (\ref{Sb3}). We end up with the following result
\begin{align}
\gamma^{(2)}(2s) \prod_{i,j=1}^2 \gamma^{(2)}(a_i+b_j)  
\int_{-\textup{i} \infty}^{\textup{i} \infty}  \prod_{i=3}^4 \gamma^{(2)}(a_i-s - z) \gamma^{(2)}(b_i-s + z)   \nonumber \\
\times \prod_{i=1}^2 \gamma^{(2)}(s+z+b_i)\prod_{i=1}^2 \gamma^{(2)}(a_i+s-z) \frac{dz}{\textup{i} \sqrt{\omega_1\omega_2}} \:.
\end{align}
Then integrating (\ref{doubleint}) over the $z$ variable one finds that
\begin{align}
\gamma^{(2)}(2s) \prod_{i,j=3}^4 \gamma^{(2)}(a_i+b_j-2s)  
\int_{-\textup{i} \infty}^{\textup{i} \infty}  \prod_{i=1}^2 \gamma^{(2)}(a_i - x) \gamma^{(2)}(b_i + x)
\nonumber \\ \times
\prod_{i=3}^4 \gamma^{(2)}(b_i+x)\prod_{i=3}^4 \gamma^{(2)}(a_i-x) \frac{dx}{\textup{i} \sqrt{\omega_1\omega_2}} \:.
\end{align}
The latter two expressions are results of the same integral expression, therefore we find that
\begin{align}
\gamma^{(2)}(2s) \prod_{i,j=1}^2 \gamma^{(2)}(a_i+b_j)  
\int_{-\textup{i} \infty}^{\textup{i} \infty}  \prod_{i=3}^4 \gamma^{(2)}(a_i-s - z) \gamma^{(2)}(b_i-s + z)\nonumber \\  
\times 
\prod_{i=1}^2 \gamma^{(2)}(s+z+b_i)\prod_{i=1}^2 \gamma^{(2)}(a_i+s-z) \frac{dz}{\textup{i} \sqrt{\omega_1\omega_2}} 
\nonumber \\ =
\gamma^{(2)}(2s) \prod_{i,j=3}^4 \gamma^{(2)}(a_i+b_j-2s)  
\int_{-\textup{i} \infty}^{\textup{i} \infty}  \prod_{i=1}^2 \gamma^{(2)}(a_i - x) \gamma^{(2)}(b_i + x)
\nonumber \\  \times 
\prod_{i=3}^4 \gamma^{(2)}(b_i+x)\prod_{i=3}^4 \gamma^{(2)}(a_i-x) \frac{dx}{\textup{i} \sqrt{\omega_1\omega_2}} \:.
\end{align}
One can rewrite this expression in a compact form
\begin{align}
\int_{-\textup{i} \infty}^{\textup{i} \infty}  \prod_{i=1}^4 \gamma^{(2)}(a_i - x) \gamma^{(2)}(b_i + x)
\frac{dx}{\textup{i} \sqrt{\omega_1\omega_2}}
\nonumber \\ =
\frac{\prod_{i,j=1}^2 \gamma^{(2)}(a_i+b_j)}{\prod_{i,j=3}^4 \gamma^{(2)}(\tilde{a_i}+\tilde{b_j})}  
\int_{-\textup{i} \infty}^{\textup{i} \infty}  \prod_{i=1}^4 \gamma^{(2)}(\tilde{a_i} - z) \gamma^{(2)}(\tilde{b_i} + z)
\frac{dz}{\textup{i} \sqrt{\omega_1\omega_2}} \;,
\end{align}
with the related balancing condition $\sum_{i=1}^4a_i+b_i= 2(\omega_1+\omega_2)$ by introducing new variables
\begin{align}
\begin{aligned}
     \tilde{a}_i & =  a_i+s, & \tilde{b}_i & =  b_i+s, & \text{if} \;\;\; i=1,2\;,
     \\ 
     \tilde{a}_i & =  a_i-s, & \tilde{b}_i & =b_i-s, & \ \text{if} \;\;\; i=3,4\;,
\end{aligned}
\end{align}
where
        \begin{align}
        \begin{aligned}
             s=\frac{1}{2}(\omega_1+\omega_2-a_1-a_2-b_1-b_2)=\frac{1}{2}(-\omega_1-\omega_2+a_3+a_4+b_3+b_4) \;.
        \end{aligned}
    \end{align}
In order to obtain the final result one needs to use the reflection property (\ref{reflectionp}) for the hyperbolic gamma function.




\begin{thebibliography}{10}
	
	\bibitem{Spiridonov:2010em}
	V.~P. Spiridonov, ``{Elliptic beta integrals and solvable models of statistical
		mechanics},'' {\em Contemp. Math.} {\bf 563} (2012)  181--211,
	\href{http://arxiv.org/abs/1011.3798}{{\tt arXiv:1011.3798 [hep-th]}}.
	
	\bibitem{Yamazaki:2012cp}
	M.~Yamazaki, ``{Quivers, YBE and 3-manifolds},''
	\href{http://dx.doi.org/10.1007/JHEP05(2012)147}{{\em JHEP} {\bf 05} (2012)
		147},
	\href{http://arxiv.org/abs/1203.5784}{{\tt arXiv:1203.5784 [hep-th]}}.
	
	\bibitem{Gahramanov:2015cva}
	I.~Gahramanov and V.~P. Spiridonov, ``{The star-triangle relation and 3d
		superconformal indices},''
	\href{http://dx.doi.org/10.1007/JHEP08(2015)040}{{\em JHEP} {\bf 08} (2015)
		040},
	\href{http://arxiv.org/abs/1505.00765}{{\tt arXiv:1505.00765 [hep-th]}}.
	
	\bibitem{Kels:2015bda}
	A.~P. Kels, ``{New solutions of the star–triangle relation with discrete and
		continuous spin variables},''
	\href{http://dx.doi.org/10.1088/1751-8113/48/43/435201}{{\em J. Phys.} {\bf
			A48} (2015) no.~43, 435201},
	\href{http://arxiv.org/abs/1504.07074}{{\tt arXiv:1504.07074 [math-ph]}}.
	
	\bibitem{Kels:2017vbc}
	A.~P. Kels and M.~Yamazaki, ``{Lens elliptic gamma function solution of the
		Yang–Baxter equation at roots of unity},''
	\href{http://dx.doi.org/10.1088/1742-5468/aaa8fd}{{\em J. Stat. Mech.} {\bf
			1802} (2018) no.~2, 023108},
	\href{http://arxiv.org/abs/1709.07148}{{\tt arXiv:1709.07148 [math-ph]}}.
	
	\bibitem{Yamazaki:2013nra}
	M.~Yamazaki, ``{New Integrable Models from the Gauge/YBE Correspondence},''
	\href{http://dx.doi.org/10.1007/s10955-013-0884-8}{{\em J. Statist. Phys.}
		{\bf 154} (2014)  895},
	\href{http://arxiv.org/abs/1307.1128}{{\tt arXiv:1307.1128 [hep-th]}}.
	
	\bibitem{Yamazaki:2015voa}
	M.~Yamazaki and W.~Yan, ``{Integrability from 2d ${\mathcal{N}}=(2,2)$
		dualities},'' \href{http://dx.doi.org/10.1088/1751-8113/48/39/394001}{{\em J.
			Phys.} {\bf A48} (2015)  394001},
	\href{http://arxiv.org/abs/1504.05540}{{\tt arXiv:1504.05540 [hep-th]}}.
	
	\bibitem{Gahramanov:2017idz}
	I.~Gahramanov and S.~Jafarzade, ``{Comments on the multi-spin solution to the
		Yang-Baxter equation and basic hypergeometric sum/integral identity},''
	\href{http://dx.doi.org/10.1142/S0217732319501402}{{\em {Mod. Phys. Lett.}}
		{\bf A34,no.18,1950140} (2019)  },
	\href{http://arxiv.org/abs/1710.09106}{{\tt arXiv:1710.09106 [math-ph]}}.
	
	\bibitem{Gahramanov:2017ysd}
	I.~Gahramanov and S.~Jafarzade, ``{Integrable lattice spin models from
		supersymmetric dualities},''
	\href{http://dx.doi.org/10.1134/S1547477118060079}{{\em Phys. Part. Nucl.
			Lett.} {\bf 15} (2018) no.~6, 650--667},
	\href{http://arxiv.org/abs/1712.09651}{{\tt arXiv:1712.09651 [math-ph]}}.
	
	\bibitem{Yamazaki:2018xbx}
	M.~Yamazaki, ``{Integrability as Duality: the Gauge/YBE Correspondence},''
	\href{http://arxiv.org/abs/1808.04374}{{\tt arXiv:1808.04374 [hep-th]}}.
	
	\bibitem{Gahramanov:2015tta}
	I.~Gahramanov, ``{Mathematical structures behind supersymmetric dualities},''
	\href{http://dx.doi.org/10.5817/AM2015-5-273}{{\em Archivum Math.} {\bf 51}
		(2015)  273--286},
	\href{http://arxiv.org/abs/1505.05656}{{\tt arXiv:1505.05656 [math-ph]}}.
	
	\bibitem{Spiridonov:2019kto}
	V.~P. Spiridonov, ``{Superconformal Indices, Seiberg Dualities and Special
		Functions},'' \href{http://dx.doi.org/10.1134/S1063779620040681}{{\em Phys.
			Part. Nucl.} {\bf 51} (2020) no.~4, 508--513},
	\href{http://arxiv.org/abs/1912.11514}{{\tt arXiv:1912.11514 [hep-th]}}.
	
	\bibitem{Tachikawa:2017byo}
	Y.~Tachikawa, ``{On 'categories' of quantum field theories},'' in {\em
		{International Congress of Mathematicians}}.
	\newblock 12, 2017.
	\newblock \href{http://arxiv.org/abs/1712.09456}{{\tt arXiv:1712.09456
			[math-ph]}}.
	
	\bibitem{Spiridonov:2009za}
	V.~P. Spiridonov and G.~S. Vartanov, ``{Elliptic Hypergeometry of
		Supersymmetric Dualities},''
	\href{http://dx.doi.org/10.1007/s00220-011-1218-9}{{\em Commun. Math. Phys.}
		{\bf 304} (2011)  797--874}, \href{http://arxiv.org/abs/0910.5944}{{\tt
			arXiv:0910.5944 [hep-th]}}.
	
	\bibitem{Spiridonov2014}
	V.~P. Spiridonov and G.~S. Vartanov, ``Elliptic hypergeometry of supersymmetric
	dualities ii. orthogonal groups, knots, and vortices,''
	\href{http://dx.doi.org/10.1007/s00220-013-1861-4}{{\em Communications in
			Mathematical Physics} {\bf 325} (2014) no.~2, 421--486},
	\href{http://arxiv.org/abs/1107.5788}{{\tt arXiv:1107.5788 [hep-th]}}.
	
	\bibitem{Krattenthaler:2011da}
	C.~Krattenthaler, V.~P. Spiridonov, and G.~S. Vartanov, ``{Superconformal
		indices of three-dimensional theories related by mirror symmetry},''
	\href{http://dx.doi.org/10.1007/JHEP06(2011)008}{{\em JHEP} {\bf 06} (2011)
		008},
	\href{http://arxiv.org/abs/1103.4075}{{\tt arXiv:1103.4075 [hep-th]}}.
	
	\bibitem{Gahramanov:gka}
	I.~B. Gahramanov and G.~S. Vartanov,
	\href{http://dx.doi.org/10.1142/9789814449243_0076}{``{Superconformal indices
			and partition functions for supersymmetric field theories},''} in {\em
		{XVIIth Intern. Cong. Math. Phys. 695-703 (2013)}}.
	\newblock 2013.
	\newblock
	\href{http://arxiv.org/abs/1310.8507}{{\tt arXiv:1310.8507 [hep-th]}}.
	\newblock
	
	\bibitem{Dolan:2011rp}
	F.~A.~H. Dolan, V.~P. Spiridonov, and G.~S. Vartanov, ``{From 4d superconformal
		indices to 3d partition functions},''
	\href{http://dx.doi.org/10.1016/j.physletb.2011.09.007}{{\em Phys. Lett.}
		{\bf B704} (2011)  234--241},
	\href{http://arxiv.org/abs/1104.1787}{{\tt arXiv:1104.1787 [hep-th]}}.
	
	\bibitem{Bozkurt:2020gyy}
	D.~N. Bozkurt, I.~Gahramanov, and M.~Mullahasanoglu, ``{Lens partition
		function, pentagon identity, and star-triangle relation},''
	\href{http://dx.doi.org/10.1103/PhysRevD.103.126013}{{\em Phys. Rev. D} {\bf
			103} (2021) no.~12, 126013}, \href{http://arxiv.org/abs/2009.14198}{{\tt
			arXiv:2009.14198 [hep-th]}}.
	
	\bibitem{Baxter:1997tn}
	R.~J. Baxter, ``{Star-triangle and star-star relations in statistical
		mechanics},''
	\href{http://dx.doi.org/10.1142/S0217979297000058}{{\em Int. J. Mod. Phys.}
		{\bf B11} (1997)  27--37}.
	
	\bibitem{Bazhanov:2007vg}
	V.~V. Bazhanov, V.~V. Mangazeev, and S.~M. Sergeev, ``{Exact solution of the
		Faddeev-Volkov model},''
	\href{http://dx.doi.org/10.1016/j.physleta.2007.10.053}{{\em Phys. Lett.}
		{\bf A372} (2008)  1547--1550},
	\href{http://arxiv.org/abs/0706.3077}{{\tt arXiv:0706.3077
			[cond-mat.stat-mech]}}.
	
	\bibitem{Gahramanov:2013rda}
	I.~Gahramanov and H.~Rosengren, ``{A new pentagon identity for the tetrahedron
		index},'' \href{http://dx.doi.org/10.1007/JHEP11(2013)128}{{\em JHEP} {\bf
			11} (2013)  128},
	\href{http://arxiv.org/abs/1309.2195}{{\tt arXiv:1309.2195 [hep-th]}}.
	
	\bibitem{Gahramanov:2014ona}
	I.~Gahramanov and H.~Rosengren, ``{Integral pentagon relations for 3d
		superconformal indices},'' \href{http://arxiv.org/abs/1412.2926}{{\tt
			arXiv:1412.2926 [hep-th]}}.
	[Proc. Symp. Pure Math.93,165(2016)].
	
	\bibitem{baxter1980fundamental}
	R.~Baxter, ``Fundamental problems in statistical mechanics v,'' {\em
		Proceedings of the 1980 Enschede Summer School} (1980)  .
	
	\bibitem{baxter2016exactly}
	R.~J. Baxter, {\em Exactly solved models in statistical mechanics}.
	\newblock Elsevier, 2016.
	
	\bibitem{baxter:1997ssr}
	R.~J. Baxter, ``Star-triangle and star-star relations in statistical
	mechanics,'' \href{http://dx.doi.org/10.1142/S0217979297000058}{{\em
			International Journal of Modern Physics B} {\bf 11} (1997) no.~01n02,
		27--37}.
	
	\bibitem{Hosomichi:2014hja}
	K.~Hosomichi, ``{A review on SUSY gauge theories on $S^3$},''
	\href{http://arxiv.org/abs/1412.7128}{{\tt arXiv:1412.7128 [hep-th]}}.
	
	\bibitem{Willett:2016adv}
	B.~Willett, ``{Localization on three-dimensional manifolds},''
	\href{http://dx.doi.org/10.1088/1751-8121/aa612f}{{\em J. Phys. A} {\bf 50}
		(2017) no.~44, 443006}, \href{http://arxiv.org/abs/1608.02958}{{\tt
			arXiv:1608.02958 [hep-th]}}.
		
	\bibitem{Bozkurt:2020gyy}
	D.~N.~Bozkurt, I.~Gahramanov and M.~Mullahasanoglu,
	``{Lens partition function, pentagon identity, and star-triangle relation},''\href{https://doi.org/10.1103/PhysRevD.103.126013}{Phys. Rev. D \textbf{103} (2021) no.12, 126013}
	\href{http://arxiv.org/abs/2009.14198}{
	\textit{[arXiv:2009.14198 [hep-th]]}}.	
	
	\bibitem{Pestun:2007rz}
	V.~Pestun, ``{Localization of gauge theory on a four-sphere and supersymmetric
		Wilson loops},'' \href{http://dx.doi.org/10.1007/s00220-012-1485-0}{{\em
			Commun. Math. Phys.} {\bf 313} (2012)  71--129},
	\href{http://arxiv.org/abs/0712.2824}{{\tt arXiv:0712.2824 [hep-th]}}.
	
	\bibitem{Kapustin:2010xq}
	A.~Kapustin, B.~Willett, and I.~Yaakov, ``{Nonperturbative Tests of
		Three-Dimensional Dualities},''
	\href{http://dx.doi.org/10.1007/JHEP10(2010)013}{{\em JHEP} {\bf 10} (2010)
		013},
	\href{http://arxiv.org/abs/1003.5694}{{\tt arXiv:1003.5694 [hep-th]}}.
	
	\bibitem{Imamura:2012rq}
	Y.~Imamura and D.~Yokoyama, ``{$S^3/Z_n$ partition function and dualities},''
	\href{http://dx.doi.org/10.1007/JHEP11(2012)122}{{\em JHEP} {\bf 11} (2012)
		122},
	\href{http://arxiv.org/abs/1208.1404}{{\tt arXiv:1208.1404 [hep-th]}}.
	
	\bibitem{Imamura:2011su}
	Y.~Imamura and S.~Yokoyama, ``{Index for three dimensional superconformal field
		theories with general R-charge assignments},''
	\href{http://dx.doi.org/10.1007/JHEP04(2011)007}{{\em JHEP} {\bf 1104} (2011)
		007},
	\href{http://arxiv.org/abs/1101.0557}{{\tt arXiv:1101.0557 [hep-th]}}.
	
	\bibitem{Faddeev:1995nb}
	L.~Faddeev, ``{Discrete Heisenberg-Weyl group and modular group},''
	\href{http://dx.doi.org/10.1007/BF01872779}{{\em Lett. Math. Phys.} {\bf 34}
		(1995)  249--254}, \href{http://arxiv.org/abs/hep-th/9504111}{{\tt
			arXiv:hep-th/9504111}}.
	
	\bibitem{woronowicz2000quantum}
	S.~Woronowicz, ``Quantum exponential function,''
	\href{http://dx.doi.org/10.1142/S0129055X00000344}{{\em Reviews in
			Mathematical Physics} {\bf 12} (2000) no.~06, 873--920}.
	
	\bibitem{Kashaev:2012cz}
	R.~Kashaev, F.~Luo, and G.~Vartanov, ``{A TQFT of Turaev\textendash{}Viro Type
		on Shaped Triangulations},''
	\href{http://dx.doi.org/10.1007/s00023-015-0427-8}{{\em Annales Henri
			Poincare} {\bf 17} (2016) no.~5, 1109--1143},
	\href{http://arxiv.org/abs/1210.8393}{{\tt arXiv:1210.8393 [math.QA]}}.
	
	\bibitem{BultThesis}
	F.~J. van~de Bult, {\em Hyperbolic hypergeometric functions}.
	\newblock PhD thesis, University of Amsterdam, 2007.
	
	\bibitem{Bazhanov:2007mh}
	V.~V. Bazhanov, V.~V. Mangazeev, and S.~M. Sergeev, ``{Faddeev-Volkov solution
		of the Yang-Baxter equation and discrete conformal symmetry},''
	\href{http://dx.doi.org/10.1016/j.nuclphysb.2007.05.013}{{\em Nucl. Phys.}
		{\bf B784} (2007)  234--258},
	\href{http://arxiv.org/abs/hep-th/0703041}{{\tt arXiv:hep-th/0703041
			[hep-th]}}.
	
	\bibitem{van2007properties}
	F.~J. van~de Bult, E.~M. Rains, and J.~V. Stokman, ``Properties of generalized
	univariate hypergeometric functions,'' {\em Communications in mathematical
		physics} {\bf 275} (2007) no.~1, 37--95.
	
	\bibitem{Dimofte:2012pd}
	T.~Dimofte and D.~Gaiotto, ``{An E7 Surprise},''
	\href{http://dx.doi.org/10.1007/JHEP10(2012)129}{{\em JHEP} {\bf 10} (2012)
		129}, \href{http://arxiv.org/abs/1209.1404}{{\tt arXiv:1209.1404 [hep-th]}}.
	

	\bibitem{Hadasz:2013bwa}
	L.~Hadasz, M.~Pawelkiewicz, and V.~Schomerus, ``{Self-dual Continuous Series of
		Representations for $\mathcal{U}_q(sl(2))$ and $\mathcal{U}_q(osp(1|2))$},''
	\href{http://dx.doi.org/10.1007/JHEP10(2014)091}{{\em JHEP} {\bf 10} (2014)
		091}, \href{http://arxiv.org/abs/1305.4596}{{\tt arXiv:1305.4596 [hep-th]}}.
	
	\bibitem{Gahramanov:2016wxi}
	I.~Gahramanov and H.~Rosengren, ``{Basic hypergeometry of supersymmetric
		dualities},'' \href{http://dx.doi.org/10.1016/j.nuclphysb.2016.10.004}{{\em
			Nucl. Phys.} {\bf B913} (2016)  747--768},
	\href{http://arxiv.org/abs/1606.08185}{{\tt arXiv:1606.08185 [hep-th]}}.
	
	\bibitem{Gahramanov:2016ilb}
	I.~Gahramanov and A.~P. Kels, ``{The star-triangle relation, lens partition
		function, and hypergeometric sum/integrals},''
	\href{http://dx.doi.org/10.1007/JHEP02(2017)040}{{\em JHEP} {\bf 02} (2017)
		040},
	\href{http://arxiv.org/abs/1610.09229}{{\tt arXiv:1610.09229 [math-ph]}}.
	
	\bibitem{Eren:2019ibl}
	E.~Eren, I.~Gahramanov, S.~Jafarzade and G.~Mogol,
	``{Gamma function solutions to the star-triangle equation},''
	\href{http://dx.doi.org/10.1016/j.nuclphysb.2020.115283}{Nucl. Phys. B \textbf{963} (2021), 115283}
	\href{http://arxiv.org/abs/1610.09229}{{\tt arXiv:1912.12271 [math-ph]}}.
	
\end{thebibliography}
\end{document}